\documentclass[prl,showpacs,twocolumn,aps]{revtex4}
\usepackage{graphicx}

\begin{document}
\title{Optical sum increase due to electron undressing}
\author{A. Knigavko and J.P. Carbotte}
\address{Department of Physics and Astronomy, McMaster University\\
Hamilton, Ontario, Canada, L8S 4M1}
\author{F. Marsiglio}
\address{Department of Physics, University of Alberta\\
Edmonton, Alberta, Canada, T6J 2J1}
%%%%\maketitle

\begin{abstract}
For a system with a fixed number of electrons, the total optical sum is a constant,
independent of many-body interactions, of impurity scattering and of temperature.
For a single band in a metal, such a sum rule is no longer independent of the
interactions or temperature, when the dispersion and/or finite bandwidth
is accounted for. We adopt such a model, with electrons coupled to a
single Einstein oscillator of frequency $\omega _{E}$, and
study the optical spectral weight. The optical sum
depends on both the strength of the coupling and on the characteristic
phonon frequency, $\omega_{E}$.
A hardening of $\omega_{E}$, due, for example, to a phase
transition, leads to electron undressing and translates into a decrease in
the electron kinetic energy and an increase in the total optical sum, as
observed in recent experiments in the cuprate superconductors.
\end{abstract}
\maketitle
\section{Introduction}

Recently there has been considerable interest \cite{vdmarel1}--\cite{norman2}
in the relationship between the kinetic energy of an electron system and its
optical oscillator strength sum rule. The optical sum, 
$S=\int_{0}^{+\infty}d\nu 
\mathop{\rm Re}%
\left[ \sigma (\nu )\right] ,$ 
measured in several high-T$_{c}$ cuprates for
in--plane conductivity shows noticeable temperature dependence
from room temperature all the way down to the zero temperature limit 
\cite{vdmarel1}. This
dependence is approximately $S \simeq c_0 - c_2 T^{2}$ but the proportionality
coefficient $c_2$ seems to change abruptly at the superconducting transition
temperature. If one describes the normal state by a tight binding band
\cite{hirsch2,vdmarel2} with only nearest neighbor hopping, the optical sum is directly
related to the negative of the kinetic energy. This also holds approximately
when further neighbors are included in the electron dispersion relation \cite{vdmarel2}.
The observed behavior of the optical sum \cite{vdmarel1,lobo1} has
revived discussion \cite{hirsch3} of the possibility of kinetic energy-driven
superconductivity.

In this context a more general issue of importance arises: what is the relationship
between the optical sum and essential characteristics of the electronic
system ? An understanding of the physical content of the optical sum is
achieved through the optical sum rule for a single band: 
\begin{equation}
\int_{0}^{+\infty }d\nu 
\mathop{\rm Re}%
\left[ \sigma _{\alpha \alpha }(\nu )\right] =-{\pi e^{2} \over \hbar^2}
\frac{1}{N}\sum_{k}%
\left(\frac{\partial \xi _{k}}{\partial k_{\alpha }} \right)^2
\frac{dn_{k}}{d\xi _{k}},  \label{1}
\end{equation}
where $e$ is the electronic charge, $N$ is the number of unit cells, $k$ is
electron momentum, $\xi _{k}$ is the electron dispersion, $n_{k}$ is the
probability of occupation of the state $k$ for a single spin, and
$\alpha$ is a Cartesian coordinate.
A more familiar form for the right hand side (RHS) of Eq. (\ref{1}) is 
\begin{equation}
\int_{0}^{+\infty }d\nu
\mathop{\rm Re}%
\left[ \sigma _{\alpha \alpha }(\nu )\right] = {\pi e^{2} \over \hbar^2}
\frac{1}{N}\sum_{k}%
\left( \partial ^{2}\xi _{k} \over \partial k_{\alpha }^2\right) n_{k}. 
\label{2}
\end{equation}
Eq. (\ref{1}) is obtained from Eq. (\ref{2})
by performing an integration by parts on the momentum $k_\alpha$. 
The merit in the optical sum rule in either form is that it relates 
the optical integral on the left 
to quantities that are easier to analyze. Note, however, that for
a parabolic band with infinite bandwidth only Eq. (\ref{2}) yields
the well known result ${\pi e^2 n \over 2m} \equiv \Omega_P^2/8 $ 
where $\Omega_P$ is the plasma frequency. 
This latter expression is valid for a system with interacting electrons, 
where $n$ is the electron density for {\it all} the electrons.
In this case the sum rule yields a constant,
independent of temperature, and does not give a hint of the underlying
interactions.

In practice one usually deals with a limited frequency range. Standard
optical experiments probe the conduction band, and the sum rule is to be
adapted correspondingly. A quadratic dependence of energy on wave vector is
often a model of choice for the dispersion of conduction electrons. When
combined with the infinite band approximation, it gives the electronic
density on the right hand side of Eq. (\ref{2}). However, in many cases it is
important to account for the finite width of the electronic band (for
example the quadratic dispersion definitely cannot be a good description
over the whole Brillouin zone). Then the optical sum rule, expressed for a
single band, definitely acquires an explicit temperature
dependence. Another implicit source of temperature dependence is the quasiparticle
occupation number $n_{k}$, which can be strongly affected by many-body interactions
amongst the electrons.

In this paper we wish to use a simple model to understand the properties
of the optical sum for a normal metal (governed by Fermi Liquid Theory)
when the different sources which can lead to its deviation
from a constant value are included. To this end we assume a
constant electron density of states with sharp cut-offs at the band
edges \cite{alexandrov87}--\cite{dogan1}. We also assume that the 
electrons are coupled to Einstein oscillators of frequency $\omega_{E}$.
The microscopic origin of this oscillator is not specified --- it could be
a phonon or a spin fluctuation.
A system of electrons coupled to an Einstein oscillator
provides an important example of an interacting system which is simple
enough that it can be analyzed in great detail in order to gain a qualitative
understanding of various phenomena \cite{cappelluti1},\cite{mars-b},\cite
{knigavko1}. In this model the optical sum is no longer equal to $\Omega _{p}^{2}/8$
but varies with temperature and with interaction strength with the oscillators.
We study how the strength of the coupling, denoted by $A$, as well as the
value of $\omega _{E}$ affect the value of the optical sum, and the expectation
value of the kinetic
energy of the electrons, as a function of temperature. In
particular we find that, if at some specific temperature $\omega_{E}$ 
undergoes a sudden hardening so that
the mass enlacement parameter $\lambda $ decreases,
then the total optical weight increases while the kinetic energy decreases,
as observed in experiment.

We begin in Sec. II with a brief review of the standard technique for
calculating optical conductivity within the Kubo formalism. The model
of a constant density of states with bandedge cutoffs allows us to make a clear
connection with the conductivity calculations that use the standard
"infinite bandwidth" approximation. This model is also well suited for simulating 
a tight binding dispersion for electrons in two spatial dimensions. We argue that
the self consistent treatment of the underlying equations for the electronic
self energy is important in this model. Numerical results and a discussion 
are presented in Sec. III.

\section{Formalism}

To evaluate the left hand side of the optical sum rule, Eq. (\ref{1}), we
need the frequency dependent conductivity $\sigma (\nu )$. Within linear
response theory this is obtained from the appropriate current--current
correlation function $\Pi$ \cite{mars-b} 
\begin{equation}
\sigma _{\alpha \alpha }(\nu )=\frac{i}{\nu }\Pi_{\alpha \alpha }(i\nu
_{n}\rightarrow \nu +i0^{+}),  \label{3}
\end{equation}
where $\alpha$ is a Cartesian coordinate, {\it x,y,z}. The response function $\Pi 
$ is analytically continued from bosonic Matsubara frequencies $i\nu _{n} \equiv
2 i\pi T n$
to the real axis by $i\nu _{n}\rightarrow \nu +i0^{+}$ \cite{mars1}. On the
imaginary axis $\Pi $ is given in the bubble approximation, 
in terms of the electronic Green's functions 
$G\left( k,i\omega _{m}\right) $, by the equation: 
\begin{eqnarray}
\Pi _{\alpha \alpha }(i\nu _{n}) = { 2e^{2} \over \hbar^2}
\frac{1}{N}\sum_{k\in BZ}
\left( \frac{\partial \xi_{k}}{\partial k_{x}} \right)^{2}
T\sum_{m=-\infty }^{+\infty }
\phantom{mm}
\nonumber\\
\phantom{mmmm}
G\left( k,i\omega _{m}+i\nu_{n}\right) G\left( k,i\omega _{m}\right),  
\label{4}
\end{eqnarray}
where $T$ is temperature and $i\omega_{m} \equiv i\pi T (2m-1)$ is 
the $m$-th fermionic Matsubara frequency.
The $k$ sum runs over the first Brillouin zone for the particular
band of interest. To evaluate Eq. (\ref{4}), the $k$ summation will be replaced 
by an energy integration (see below).

When the Green's functions in Eq. (\ref{4})
are expressed though the electron spectral density 
$A(k,\omega )=-\frac{1}{\pi }
\mathop{\rm Im}%
G(k,i\omega _{n}\rightarrow \omega +i0^{+})$, the formula for the real part
of the in--plane optical conductivity assumes the form 

\begin{eqnarray}
\mathop{\rm Re}\left[ \sigma _{xx}(\nu )\right] = {2\pi e^{2} \over \hbar^2}
{1 \over N} \sum_k
\left( \frac{\partial \xi_{k}}{\partial k_{x}}\right)^{2}
\int_{-\infty }^{\infty }d \omega
\phantom{mmmm} 
\nonumber\\
\phantom{mmmm}
A\left( \xi_k,\omega\right)\,A\left( \xi_k,\omega+\nu\right) 
\frac{f_{F}\left( \omega\right) -f_{F}(\omega+\nu)}{\nu},  
\label{5}
\end{eqnarray}
where $f_{F}(\omega )$ is the Fermi--Dirac distribution, and $A\left(\xi_k,\omega\right)$
is the electron spectral function.

On the RHS of the optical sum rule, Eq. (\ref{1}), the particle occupation
number $n_{k}=n(\xi_k)$ is also expressed though the Green's function: 
\begin{equation}
n(\xi_k) = \int_{-\infty }^{+\infty }d\omega\,A\left( \xi_k,\omega\right) 
f_{F}\left(\omega\right)  \label{6}
\end{equation}
and the RHS can be written as

\begin{equation}
{\rm RHS}=- {\pi e^2 \over \hbar^2} \int_{-\infty }^{+\infty }d\omega f_{F}\left(
\omega\right) {1 \over N} \sum_k\left( \frac{\partial
\xi_{k}}{\partial k_{x}}\right)^{2} \frac{\partial A\left( \xi_k,
\omega\right) }{\partial \xi_k}.  \label{7}
\end{equation}

Eq. (\ref{7}) is closer to the starting point of the sum rule derivation,
in that the same factor of the Fermi velocity squared, $\left( \partial
\xi_{k}/\partial k_{x}\right) ^{2},$ occurs in both equations (\ref{5}) and 
(\ref{7}); thus if an approximate band structure
is introduced at this step the sum rule will hold exactly.

Here we will consider two possible choices for the band structure and hence
group velocity. In a model with quadratic dispersion 
with lower band edge at $\xi = -W/2$ we have 
$\left( {1 \over \hbar}\frac{\partial \xi_{k}}{\partial k_{x}}\right)^{2}=
{2 \over mD} \left( W/2 + \xi \right)$,
where $D$ is
the dimensionality and $m$ the free electron mass. As is usual, we also
adopt a constant density of states, with $g(\xi) = 1/W$ for $-W/2 < \xi < W/2$.
Here $W$ is the bandwith, and the density of states obeys the usual sum rule.
An integration by parts,
assuming this constant density of states, then leads to the result 
\begin{equation}
{\rm RHS}= {2 \over D} {\pi e^2 n\over 2 m}
\left[ 1 -{2 \over n}\int_{-\infty}^{\infty}d \omega\,f_F(\omega) 
A(W/2,\omega)
\right],  \label{8}
\end{equation}
where $n$ is the electron density in the band.
Note that the result is now temperature dependent, and dependent on interactions.
In this expression the electron spectral density $A(\xi,\omega)$ is to 
be evaluated at the unperturbed band edge $\xi = W/2$. Thus, in the limit
of large bandwidth the second term goes to zero, and we are left with a constant
result which is within a factor of $2/D$ of the usual sum rule in three dimensions.
A precise agreement is in general not expected, since, for a single band the sum
rule will depend on the details of the dispersion, etc.

For a tight binding band the group velocity depends on wavevector in an
essential way. One can introduce a weighted density of states \cite{mars2},
$g_{xx}(\xi) \equiv {1 \over N} \sum_k (\partial \xi_k/\partial k_x)^2 \delta
(\xi - \xi_k)$, so that the Brillouin zone sum in Eq.(\ref{7}) is reduced
to a single energy integration. 
If only the nearest neighbor hopping is included, in one dimension one finds 
that $g_{xx}(\xi)/g(\xi)= (2ta)^2 [1 - (\xi/2t)^2]$, 
where  $a$ is the lattice spacing and $t$ is the single particle hopping. 
Here $g(\xi)$ is the single electron density of states.
In higher dimensions one can obtain somewhat more complicated
expressions involving complete elliptic integrals of the first and second kind,
but the approximation
$g_{xx}(\xi)/g(\xi) \approx \left(Wa/(2D)\right)^2
\left[1 - \left({2 \xi / W}\right)^2\right]$
remains excellent, particularly near the band edges. 
Note that in $D$ dimensions the hopping integral can be expressed through 
the bandwidth as $t=W/(4D)$.
Thus it becomes natural to use the replacement 
\begin{equation}
\left( {\partial \xi_{k} \over \partial k_{x}}\right) ^{2} =
{W\over D} {\hbar^2 \over 2m_b}
\left[1 - \left({ \xi \over W/2}\right)^2\right], 
\label{9}
\end{equation}
where we introduced the mass $m_b$ of an electron in a tight binding band
using the standard definition ${\hbar^2 \over 2m_b} = ta^2$.
Substituting this expression into Eq. (\ref{7}), we obtain
\begin{eqnarray}
{\rm RHS} = {2\over D}{\pi e^2 \over 2m_b} \biggl[
- \int_{-\infty}^{+\infty}{d\omega \over W/2} \, 
\phantom{mmmmmmmm}
\nonumber\\[0.03in]
\phantom{mmmmmmmm}
f_F(\omega)\int_{-W/2}^{+W/2} {d\xi \over W/2}  \, \xi A(\xi,\omega)
\biggr] \label{10}.
\end{eqnarray}
Note that the quantity in the square brackets in Eq. (\ref{10}) 
is just the negative of the
kinetic energy, which is a well known result \cite{hirsch2,vdmarel2} for
a tight-binding model with nearest neighbour hopping only. 

In everything that follows we will retrict ourselves to half-filling. Then the model
has particle-hole symmetry, and Eq. (\ref{10}) reduces to
\begin{eqnarray}
{\rm RHS} = {1\over D}{\pi e^2 \over 2m_b}
\biggl[ 1 - 4 \int_{-\infty}^{+\infty}{d\omega \over W/2} \, 
\phantom{mmmmmmm}
\nonumber\\[0.03in]
\phantom{mmmmmm}
f_F(\omega) 
\int_0^{W/2}{d\xi \over W/2} \, \xi A(\xi,\omega)
\biggr] \label{11}.
\end{eqnarray}
Eq. (\ref{10}) or
Eq. (\ref{11}) reduces to the usual result for large bandwidth, within a factor
of order unity involving the dimensionality, as in the case with quadratic dispersion.

To compute the conductivity given by Eq. (\ref{5}) and the optical sum rule
given by Eqs. (\ref{8}) and (\ref{11}) we require the electron
self energy $\Sigma (\omega + i \delta)$, which determines the electron Green's
function $G(\xi,\omega + i \delta)$ and spectral function $A(\xi,\omega)$. 
One possibility is to use a model for the
self energy (see the paper by Norman and Pepin \cite{norman1} where they
obtain $\Sigma$ from a fit to APRES data for example). We use a more microscopic
approach and assume that the
electrons are coupled to bosons which are modeled by Einstein oscillators. While
we adopt the formalism for the conventional electron phonon mechanism, we are
open to the possibility that the electrons interact with spin fluctuations,
and we therefore tacitly assume that this formalism applies in this case as well.
The interaction is defined in terms of the
electron-boson spectral density, $\alpha ^{2}F(\omega)$, which for
an Einstein oscillator is simply a delta function: 
$\alpha ^{2}F(\omega)=A\,\delta (\omega-\omega_E)$ where $\omega_E$ is
the Einstein frequency.
The parameter $A$ specifies the strength of the interaction (not to be confused 
with the spectral function used above); it is given by
$A = \lambda \omega_E /2$.
This quantity can be conveniently visualized
as the area under the $\alpha ^{2}F(\omega)$ curve for arbitrary (i.e. non
delta-function-like) electron--boson spectral densities. 
The parameter $\lambda $ is the usual electron mass enhancement parameter.
Two of these three parameters ($\lambda $, $\omega_E$, and $A$) are independent.

The self energy equations for $\Sigma(\omega + i \delta)
\equiv \Sigma_{1}(\omega + i \delta) + i\Sigma_{2}(\omega + i \delta)$ have the
form:
\begin{eqnarray}
\Sigma_1(\omega + i \delta) =\, A\, P\int_{-\infty }^{\infty }d\omega
^{\prime }\biggl[ \frac{f_{B}(\omega_E) + f_{F}(-\omega^{\prime })}{\omega
-\omega_E -\omega^{\prime}}
\phantom{mm}
\nonumber\\[0.03in]
\phantom{mm}
+\frac{f_{B}(\omega_E) + f_{F}(\omega ^{\prime })}
{\omega +\omega_E -\omega^{\prime }}\biggr] N(\omega^{\prime }),
\label{self-eq-1} \\[0.08in]
\Sigma_{2}(\omega + i \delta) = -A\, \pi \left[ N(\omega -\omega_E)\left\{
f_{B}(\omega_E) + f_{F}(\omega_E \right.\right. 
\nonumber \\[0.03in] 
%\phantom{mmmm}
\left. -\omega ) \right\} + 
\left. N(\omega +\omega_E)
\left\{ f_{B}(\omega_E) + f_{F}\left(\omega_E +\omega \right) \right\} \right] ,
 \label{self-eq-2} \\[0.08in]
N(\omega ) =\frac{1}{\pi }\biggl[ \tan ^{-1}\frac{\omega -\Sigma
_{1}(\omega )+ W/2}{-\Sigma _{2}(\omega )}
\phantom{mmmmm}
\nonumber\\[0.03in]
\phantom{mmm}
-\tan ^{-1}\frac{\omega -\Sigma
_{1}(\omega )- W/2}{-\Sigma _{2}(\omega )}\biggr] ,  \label{self-eq-3}
\end{eqnarray}
where the symbol $P$ in Eq. (\ref{self-eq-1}) denotes the Cauchy principal
value of the integral and $f_{B}(\omega_E)$ is the Bose--Einstein
distribution function. The form of Eq. (\ref{self-eq-3}) is a consequence of the 
model for the electronic band we have adopted.

For an infinite quadratic band the self energy is an explicit function of
frequency \cite{mars1}, which is given by Eqs. (\ref{self-eq-1})--(\ref{self-eq-2}) 
with $N(\omega)=1$ according to Eqs. (\ref{self-eq-3}). However, if one tried to keep 
using Eqs. (\ref{self-eq-1})--(\ref{self-eq-2}) with $N(\omega)=const$ to compute 
the (dimensionless) renormalized density of states 
$N(\omega) = W \int d \xi g(\xi) \,A\left(\xi,\omega\right) $
%= \int_{-W/2}^{+W/2}d \xi\,A\left(\xi,\omega\right)$ 
for a band of  finite width, then no satisfactory result could be produced.
In this case it is necessary to solve Eqs. (\ref{self-eq-1})--(\ref{self-eq-3}) for 
the self energy $\Sigma (\omega + i \delta)$ self-consistently \cite{mitrovic1}.
The importance of this is illustrated in Fig. 1. 
Here we show both self-consistent (solid curve) and non self-consistent 
(dashed curve) results for $N(\omega)$ in the electronic band at reduced 
temperature $t \equiv T/(W/2) = 0.02$ for $\Omega \equiv \omega_E/(W/2) = 0.1$ 
and $\lambda =2$. 
Note that the non self-consistent electron density of states (dashed curve)
shows an unphysical saturation for a small range of frequencies above the bare band 
edge (at a value of $\omega/(W/2) = 1$). On the other hand, the
self-consistent density of states gradually decreases with increasing $\omega $
over a range of frequencies equal to a fraction of the bandwidth.
Further details of this model
will be provided elsewhere \cite{note}. We also refer the reader to
papers \cite{alexandrov87}--\cite{dogan1}\ in which coupling of the electrons
to phonons is considered within a Migdal-Eliashberg self-consistent approximation.

Before proceeding to the presentation of numerical results for the optical
integral, note that there are two ways to calculate it. The easier way is
through direct evaluation of Eq. (\ref{8}) or (\ref{11}), which is the RHS of
the optical integral sum rule. The harder way is to evaluate the conductivity
(Eq. (\ref {5})) and then integrate it explicitly over all frequencies.
This latter method gives us an
understanding of how the optical spectral weight is distributed in
frequency \cite{note}. We have used both methods, and find agreement
with an accuracy of $0.01\%$.

%%% Fig.1 
\begin{figure}[tp]
\begin{center}
\includegraphics[height=5.cm]{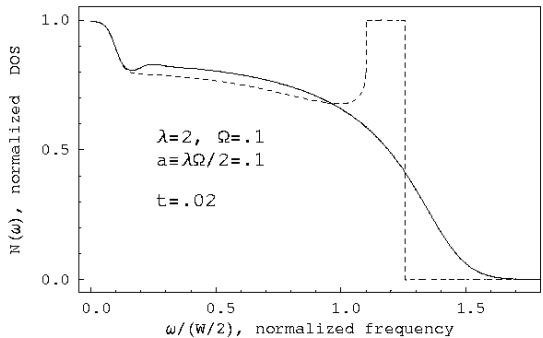}
\caption{
Normalized density of states $N(\omega )$ vs normalized frequency
$\omega/(W/2)$ in an electronic band renormalized by the interaction
with Einstein oscillator of normalized frequency $\Omega =0.1$. 
Shown are the results of self-consistent (solid curve) and non self-consistent 
(dashed curve) calculations for the electronic self energy. The mass enhancement 
parameter is $\lambda =2$, normalized temperature is $t=0.02$.
}
\end{center}
\end{figure}

%%% Fig. 2
\begin{figure}[th]
\begin{center}
\includegraphics[height=10cm]{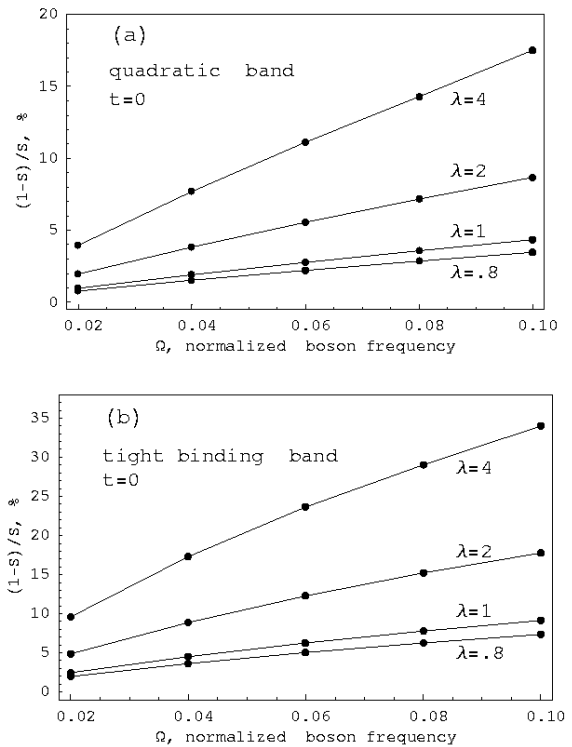}
\caption{
Percent deviation from $1$ of the optical sum as a function of
normalized boson frequency $\Omega $ for several values of the mass enhancement
parameter $\lambda$. 
The optical sum deviation from 1 is negative in all cases although here 
we quote the percentage as a positive quantity.
Results are
shown for (a) quadratic dispersion and (b) tight binding dispersion. The
curves in panel (b) also represent the change in the kinetic energy, irrespective
of the above choice of the band structure.
}
\end{center}
\end{figure}

\section{Discussion of the results}

For purposes of presentation, we show results for the optical integral $S$ 
in dimensionless form, i.e. by omitting the factor that precedes the square brackets
in Eq. (\ref{8}) for the quadratic dispersion, and in Eq. (\ref{11}) for the
tight-binding dispersion. Thus, the `standard' value for the sum rule in the
ensuing results corresponds to a value 1.
All energies are measured in units of $W/2$, half of the bare electronic bandwidth.
We also use normalized variables: the normalized frequency of Einstein oscillators, 
$\Omega \equiv \omega_E/(W/2)$, the normalized temperature, $t \equiv T/(W/2)$, 
and the normalized area under $\alpha^2F(\omega)$, $a \equiv A/(W/2)$.

In Fig. 2 we show zero temperature results for the percent deviation of the
total optical spectral weight which results from interactions, i.e. from a
finite value of the parameter $a$ which enters the electron boson spectral
density. In all cases the deviations are negative but they are plotted as
positive as a function of the normalized boson frequency, $\Omega$. The top 
(bottom) frame applies to the case with quadratic (tight-binding) dispersion.
The various curves are labeled by the value of
electron mass enhancement parameter $\lambda $ that were used.  We see that
for a fixed value of $\lambda$ the percent deviation increases as $\omega_E$
increases; similarly, for fixed $\omega_E$ and increasing $\lambda$ the deviation
increases.
While both band structure
models show the same qualitative behavior the effect is larger for the tight
binding case. Note that the result in the lower frame also represents 
the kinetic energy change.
The optical sum and the kinetic energy do follow each other but are
numerically different for the quadratic band case.

In Fig. 3 we consider temperature variations of the optical sum as a
function of normalized temperature $t$. Results for $a=0.02$ (top frame) and  
$a=0.1$ (bottom frame) are shown. The former value corresponds to conventional 
metals while the latter number is characteristic of strongly coupled systems 
like the high--T$_{c}$ cuprates. 
For Fig. 3 we have chosen two representative values of the boson frequency
$\omega_E$: in the top frame they correspond to $\lambda =0.8$ and $1$, 
while in the bottom frame they correspond to $\lambda =2$ and $4$.
In both frames we show the optical sum derived from both the quadratic band and the
tight binding band, as indicated. In each frame solid and dashed curves are 
used to distingush between the results for two different values of the electron 
enhancement parameter. 
We see that, with $a$ and $\lambda$ parameters fixed, the  temperature
variations of the optical sum for tight binding are larger than for quadratic bands. 

%%% Fig. 3
\begin{figure}[th]
\begin{center}
\includegraphics[height=10cm]{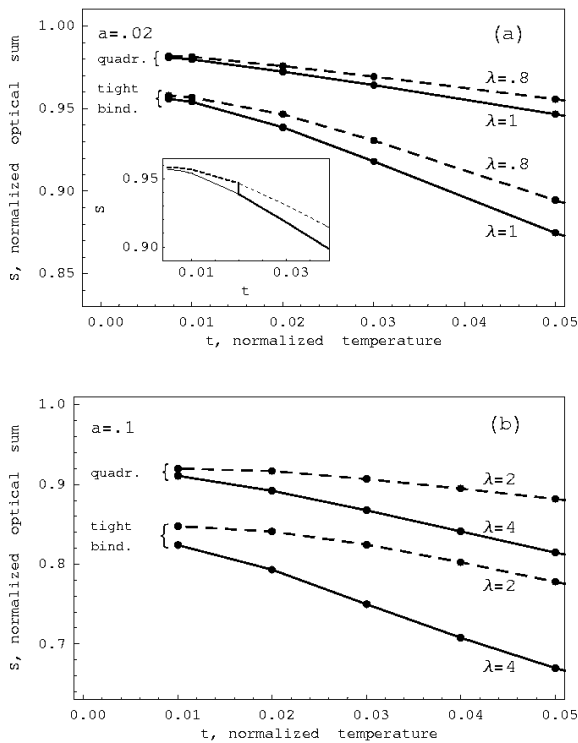}
\caption{
The variation of the optical sum vs. temperature for the interaction
strengths (a) $a=0.02$ and (b) $a=0.1$. The results are for both the quadratic
and tight binding band, as indicated. In panel (a) the mass enhancement
parameter $\lambda =0.8$ (dashed curves) and $\lambda =1$ (solid curves), while 
in panel (b) $\lambda =2$ (dashed curves) and $\lambda =4$ (solid curves).
The insert in panel (a) illustrates the behavior of the optical sum
during a sudden ``undressing transition'' at $t_{undress}=0.02$ with 20\%
hardening of the normalized boson frequency $\Omega $ and a corresponding reduction of
the the mass enhancement parameter from $\lambda =1.0$ \ to $\lambda =0.8.$
}
\end{center}
\end{figure}

In all cases the variation of the optical sum with temperature 
increases as $\lambda$ increases. 
At the same time the absolute value of the deviation of $S$ from 1 at $t=0$ increases 
with increasing coupling strength $a$ (compare the top frame with the bottom frame).
This is in accordance with the results presented in Fig. 2
where we see that for a fixed $\lambda$ the absolute value of the deviation grows 
as $\Omega$ increases (remember that $a=\lambda \Omega /2$). 

From Fig. 3 we conclude 
that interactions play an essential role in determining the temperature
dependence of the optical sum. To understand this point better we return to Eq.
(\ref{2}) which gives the optical sum as an integral of two factors,
the second derivative of the electron dispersion, and the occupation probability,
$ n(\xi_{k})$, given by Eq. (\ref{6}). 
The temperature dependence of $ n(\xi_{k})$ derives from two sources: the
Fermi function $f_{F}\left(\omega\right)$ and the electron spectral function 
$A\left(\xi,\omega\right)$.
The former factor is always operative, even in the noninteracting case. 
The latter factor produces an additional temperature dependence only when self
energy effects are included (check Eqs. (\ref{self-eq-1})--(\ref{self-eq-2}) 
which include temperature through
the functions $f_{F}\left(\omega\right)$ and $f_{B}\left(\omega\right)$).

To get some idea of the significance of this second source of the temperature 
dependence  we evaluate Eq. (\ref{6}) with the thermal factor 
$f_{F}\left(\omega\right)$ artificially "switched off" and kept in the form 
valid at $t=0$ (i. e. in the form of the step function), but with the spectral 
function $A\left(\xi,\omega\right)$ evaluated properly from 
Eqs. (\ref{self-eq-1})--(\ref{self-eq-2}) for a range of temperatures.
The results are shown in Fig. 4a for $t=0.0, 0.01, 0.03$ and $0.05$. Sharper curves
correspond, of course, to lower temperatures. We see that the temperature dependence 
of $ n(\xi)$ obtained in this way is quite strong. 

The full temperature dependence of the occupation probability $ n(\xi)$, resulting 
when both the thermal factor $f_{F}\left(\omega\right)$ and the temperature
dependence of $A\left(\xi,\omega\right)$ are accounted for in Eq. (\ref{6}),
is illustrated in Fig. 4b by dashed curves. We have also included the
corresponding results 
from Fig. 4a so a direct comparison could be made. The conclusion is that
the temperature dependence of the self energy is always important for determining the
occupation probability $n_{k}(T)$ and therefore plays an essential
role in the sum rule. This source of the temperature dependence of $S$ becomes 
dominant as the temperature increases. This important
dependence is omitted in Ref. \cite{vdmarel2}; in that analysis interactions were
not included.

%%% Fig. 4
\begin{figure}[th]
\begin{center}
\includegraphics[height=10cm]{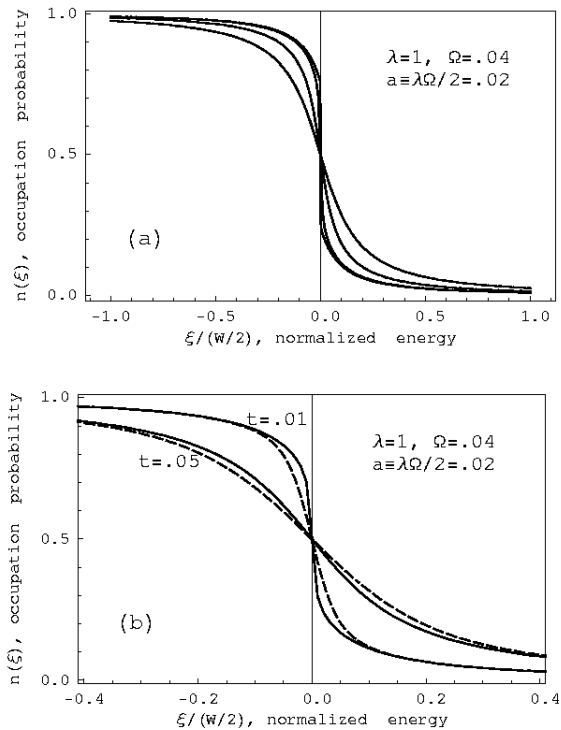}
\caption{
The probability of occupation of the state $n(\xi)$ vs normalized
energy $\xi/(W/2)$.
(a) The results for the case when only the
temperature dependence of the self energy is included. The normalized 
temperatures  are $t=0.0, 0.01, 0.03$ and $0.05$, with broader curves corresponding 
to higher temperatures.
(b) Comparison of the results when the complete temperature dependence 
of $n(\xi)$ (see Eq. (6)) is accounted for (dashed curves) with the case 
presented in (a) (solid curves). The normalized temperatures are $t=0.01$
and $0.05$, as indicated. 
}
\end{center}
\end{figure}

%%% Fig. 5
\begin{figure}[th]
\begin{center}
\includegraphics[height=10cm]{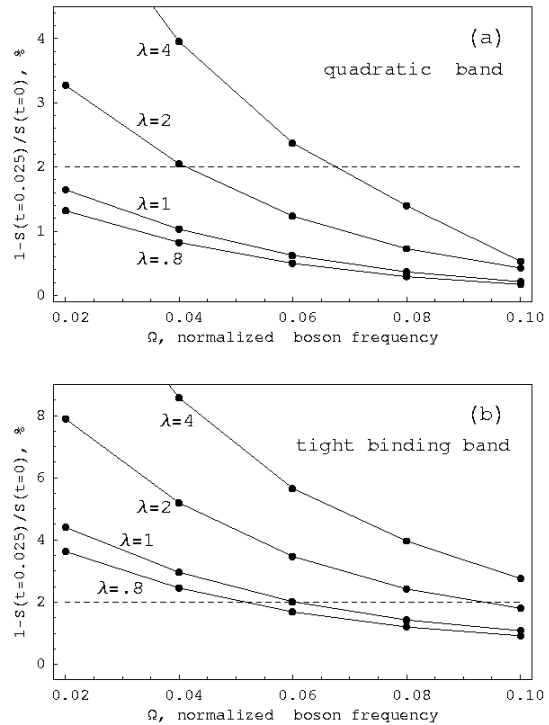}
\caption{
The percent variation in the optical sum between normalized
temperatures $t=0.025$ and $t=0$ vs. the normalized boson frequency 
$\Omega $ for the values of mass enhancement parameter $\lambda =0.8 ,1, 2$,
and $4$. 
Panel (a) applies to the parabolic band, and panel (b) to the tight binding band. 
The dashed line shows the 2\% loss of $S$ estimated from the data of 
Molegraaf {\it et al} \cite{vdmarel1}.
}
\end{center}
\end{figure}

In Fig. 5 we show the percentage spectral weight lost between $t=0$ and 
$t=0.025$ as a function of the normalized boson energy $\Omega $ for several values 
of the mass enhancement parameter $\lambda$. 
We see that the percentage increases
with increasing $\lambda$ but that for a fixed $\lambda $ it decreases with
increasing values of $\omega_E $ in the range of parameters considered. 
In a recent experiment Molegraaf {\it et al} \cite{vdmarel1} have found 
that the optical sum decreases noticeably from $0$ K to $200$ K in two cuprate
superconducting samples. From the data in this reference we estimate the 
corresponding percentage change to be of the order of 2\%.
We show this as a horizontal dotted line in Fig. 5. 
Note that in these experiments the bandwidth $W$ is
estimated to be 1.25 eV and $T=200$ K corresponds to $t=.027$, close to the
value used in Fig. 5. While our theory is simple with coupling to a single
boson only, the experimental observation puts a constraint on allowed
values of $\Omega$ and $\lambda$. Only those results that fall close to the dotted
line are possible. This still leaves a considerable range of possible
parameters. If one has in mind a particular boson model such as phonons or
spin fluctuations, $\omega_E $ is further constrained and the optical sum rule
can be used to deduce the value of $\lambda$. Alternatively, from a
detailed fit to the frequency dependence of the optical conductivity in the
cuprates, Schachinger and Carbotte \cite{carbotte-book} have
determined that $\lambda \simeq 2$ in these materials. Reference to the
lower frame of Fig. 5 gives an estimate of $\Omega \approx 0.1$,
This implies a frequency $\omega_E \approx 62$ meV, which is somewhat high
for phonons.
Extensive calculations 
\cite{carbotte-rmp}\ for a distributed electron--boson spectra $\alpha
^{2}F(\omega )$ suggest that the appropriate single frequency characterizing
the spectrum is $\omega_{\ln }$, which is equal to approximately one half
of the maximum phonon energy ($\omega_{D}$). For the cuprates this should
be $\lesssim 40\,$ meV.

We will not pursue this point here \cite{note}, but instead focus on the
primary observation of Molegraaf and coworkers\cite{vdmarel1}. They
find an abrupt jump upwards in the optical spectral weight as the temperature is
lowered into the superconducting state in two cuprate materials. They interpret this
as indicative of a {\it decrease} in the absolute value of the kinetic energy.
This is contrary to what is expected in the conventional BCS framework and is
suggestive of a novel type of {\it kinetic energy driven} superconductivity. 
Returning to the top frame of Fig. 3 note that if at a critical
temperature $T_{c}$ a phase transition occurs in which the
boson energy hardens (leaving everything else the same) so that $\lambda $
changes from 1.0 to 0.8 say, the corresponding total optical spectral weight
would jump from the solid to the dashed line. This is illustrated in the inset.
`Undressing' of the electron's
mass due to hardening of the boson spectrum leads directly to an increase in
the optical sum and a decrease in the kinetic energy. We have not considered
specifically the superconducting transition. In this case the reduction in
kinetic energy due to the undressing process would have to overcompensate
for the increase in kinetic energy that occurs when Cooper pairs form.
Schachinger, Carbotte and Basov \cite{carbotte1} (see also reference \cite
{carbotte-book}) have determined the boson spectrum involved for electron
interactions in the cuprates from considerations of the frequency dependence of
the infrared
conductivity.
They found that, as the temperature is reduced, the
boson spectrum becomes gapped at low frequency with formation of an optical
resonance at higher frequency, a process which effectively corresponds to a
hardening of the boson spectrum; this change would manifest itself as the 
`undressing' process
described here. 
A similar conclusion was reached in Ref. \cite{haslinger03} based on calculations
of the condensation energy.

\section{Conclusions}

We have adopted a very simple model for interacting electrons to investigate
the dependence of the optical sum on interactions and temperature.
The model consists of electrons with bandwidth $W$ with a constant density of states,
and with a dispersion given by either a parabolic relation or a tight-binding
description. This additional modelling is required for the dispersion in order to
correctly describe the group velocity, whose energy dependence is important for
satisfying the optical sum rule. These electrons interact with a boson, and
we have described this interaction with a (self-consistent) Migdal approximation.
The optical conductivity is described by the bubble approximation; one can readily
verify that the optical sum rule, which relates an exact two particle response function
to an exact single particle property, is satisfied by these two (seemingly) unrelated
approximations.

A natural interpretation of the sum rule experiments 
\cite{vdmarel1,lobo1,vdmarel2,hirsch3} is to infer a novel mechanism for
superconductivity that is accompanied by a kinetic energy decrease (absolute value),
in contrast to the usual BCS case. Here we have adopted an approach which
in this sense is conventional; the deviation from the usual BCS case instead
arises because of a change in the boson characteristics at and below the
superconducting transition. This possibility was inspired by an analysis of the neutron
scattering data which showed a definite change in the spin fluctuation
spectrum at $T_c$ \cite{carbotte-book,carbotte1}. 
Here we have modelled these changes as a shift in boson spectral weight
from low to high frequency with a concomitant lowering of $\lambda$.
As our results show, this leads naturally to an {\it increase} in the optical
sum, as observed in experiment. Thus, these experiments, along with others
\cite{carbotte-book,carbotte1,timusk04} find a consistent explanation in
boson mediated superconductivity accompanied by temperature dependent changes
in the boson spectral function.

\section{Acknowledgments}

Work supported by the Natural Science and Engineering Research Council of
Canada (NSERC) and the Canadian Institute for Advanced Research (CIAR).

\end{document}